\documentclass[a4paper]{jpconf}
\usepackage{graphicx}
\def\be{\begin{eqnarray} &&} 
\def\ee{\end{eqnarray}}

\newcommand{\bq}{\begin{eqnarray}}
\newcommand{\eq}{\end{eqnarray}}

\newcommand {\vece}[1]{\overset{_\rightarrow}{#1}}

\mathchardef\bbrho="711A
\mathchardef\bbsigma="711B
\mathchardef\bbtau="711C
\mathchardef\bbvarrho="7125
\mathchardef\bbvarsigma="7126
\mathchardef\bbxi="7118

\def\be{\begin{eqnarray} &&}

\def\ee{\end{eqnarray}}
\newcommand\la{\langle}
\newcommand\ra{\rangle}

\def \brho {{\mbox{\boldmath$\rho$}}}

\def\beq{\begin{equation}}
\def\eeq{\end{equation}}
\usepackage{amsbsy,amsmath}

\begin{document}

\title{Three--Dimensional parton structure of light nuclei}



\author{Sergio Scopetta$^1$,         
        Alessio Del Dotto$^{2,3}$,
       Leonid Kaptari$^4$,
       Emanuele Pace$^5$,
        Matteo Rinaldi$^6$ and
        Giovanni Salm\`e$^2$      }


\address{$^1$
       Dipartimento di Fisica e Geologia,   
Universit\`a di Perugia and INFN, Sezione di Perugia, Italy}
\address{$^2$
       INFN, Sezione di Roma, Italy}
\address{$^3$ University of South Carolina, Columbia, SC 29208, USA}
\address{$^4$ 
         Bogoliubov Lab. Theor. Phys., 141980, JINR, Dubna, Russia  }
\address{$^5$        
      Dipartimento di Fisica,  Universit\`a di Roma Tor Vergata 
and INFN, Sezione di Roma Tor Vergata, Italy}
\address{$^6$ 
      Departamento de F\`isica Te\`orica, Universidad
de Valencia and IFIC, CSIC, Valencia, Spain}

%\maketitle

\begin{abstract}
Two promising 
directions beyond inclusive deep inelastic scattering experiments, aimed at
unveiling the three dimensional structure of the bound 
nucleon, are reviewed, considering in particular
the $^3$He nuclear target. 
The 3D structure in coordinate space can be accessed through deep 
exclusive processes, whose non-perturbative part
is encoded in
generalized parton distributions. 
In this way, the distribution of partons 
in the transverse plane can be obtained. 
As an example of a deep exclusive process,
coherent deeply virtual
Compton scattering (DVCS) off
$^3$He nuclei, important
to access the neutron generalized parton distributions (GPDs),
will be discussed.
In Impulse Approximation (IA), the sum of 
the two leading twist, quark helicity conserving
GPDs of $^3$He, $H$ and $E$, at low momentum transfer, turns out to
be dominated by the neutron contribution.
Besides, a technique, able to take into account the nuclear effects 
included in the Impulse Approximation analysis,
has been developed.
The spin dependent
GPD $\tilde H$ of $^3$He is also found to be largely dominated, 
at low momentum transfer, by the neutron contribution.
The knowledge of the GPDs $H, E$ and $\tilde H$ of $^3$He 
is relevant for the planning of
coherent DVCS off $^3$He measurements.
Semi-inclusive deep inelastic scattering processes
access the momentum space 
3D structure parameterized through transverse momentum dependent parton 
distributions. 
A distorted spin-dependent spectral function 
has been recently introduced for $^3$He, 
in a non-relativistic framework, 
to take care 
of the final state interaction between the observed pion and the remnant in  
semi-inclusive
deep inelastic electron scattering off transversely
polarized $^3$He.
The calculation of the Sivers and Collins single spin asymmetries for $^3$He,
and a straightforward
procedure to effectively
take into account   
nuclear dynamics and final state interactions,
will be reviewed.
The Light-front dynamics generalization of the analysis  
is also addressed. 

\end{abstract}

\section{Introduction}
\label{intro}

The nucleus is a unique laboratory for studies of the QCD hadron 
structure. Indeed,
nuclear targets are required, e.g., for
the extraction of the neutron information from light 
nuclei and a precise flavor separation of parton distributions 
(PDFs), for the measurement of nuclear PDFs, relevant for the analysis of 
heavy ions collisions aimed at producing quark-gluon plasma, or to study 
in-medium fragmentation, mandatory to clarify the dynamics of 
hadronization. Unfortunately, inclusive Deep Inelastic 
Scattering (DIS) off nuclei has proven to be unable to answer 
fundamental questions, such as the quantitative microscopic 
explanation of the European Muon Collaboration (EMC) 
effect~\cite{Aubert:1983xm}, i.e., the medium 
modification of PDFs, the quantitative understanding of 
the parton structure of the neutron, the study in-medium of the 
distribution of parton transverse momentum, relevant for studies of 
hadronization as well as of chiral-odd quantities, such as the transversity 
PDFs or the Sivers and Collins functions and asymmetries.
Thanks to
novel coincidence measurements, possible at high luminosity facilities
such as 
Jefferson Laboratory (JLab),  
a new era in the knowledge of the parton structure of nuclei
has started~\cite{dupre_scope}.
In particular, two promising directions beyond 
inclusive measurements, aimed at accessing the three dimensional (3D) 
structure of the bound nucleon, are deep exclusive processes off nuclei,
and semi-inclusive deep inelastic 
scattering (SIDIS) involving nuclear targets.
In deep exclusive processes, such as
deeply virtual Compton Scattering (DVCS), one accesses 
the 3D structure in coordinate space in terms
of generalized parton distributions (GPDs)~\cite{gpds_here};
in SIDIS, the momentum space 3D 
structure is obtained by studying transverse momentum dependent 
parton distributions (TMDs)~\cite{gpds_here}.
In the following, we summarize relevant results
obtained by our group, for the $^3$He target, in the last
few years, mainly with the aim of extracting the neutron
information.
The next section is devoted to the analysis of $^3$He GPDs,
entering the DVCS cross sections of experiments which could be planned
at the high-luminosity facilities;
in the third section, an analysis of SIDIS
off $^3$He, taking into account final state interactions
between the produced meson and the remnants, will be reviewed.
In the last section, perspectives are 
addressed and conclusions
are drawn.

\section{Deeply virtual Compton Scattering off $^3$He}

Initially introduced in Ref. \cite{uno},
GPDs are a crucial source of unique information.
They represent in particular a tool to shed light on 
the so called ``Spin Crisis'' problem.
GPDs measurements will
allow indeed to access the parton total angular momentum \cite{due}.  
Then, by subtracting from the latter the helicity quark contribution, 
measured in other independent processes, 
the parton orbital angular momentum (OAM)
could be estimated. 

\begin{figure*}[t]
\vspace{6.5cm}
%\special{psfile=ff_fermi.eps hoffset=0
%voffset=70
%hscale= 45 vscale= 45 
%angle=0}
\includegraphics{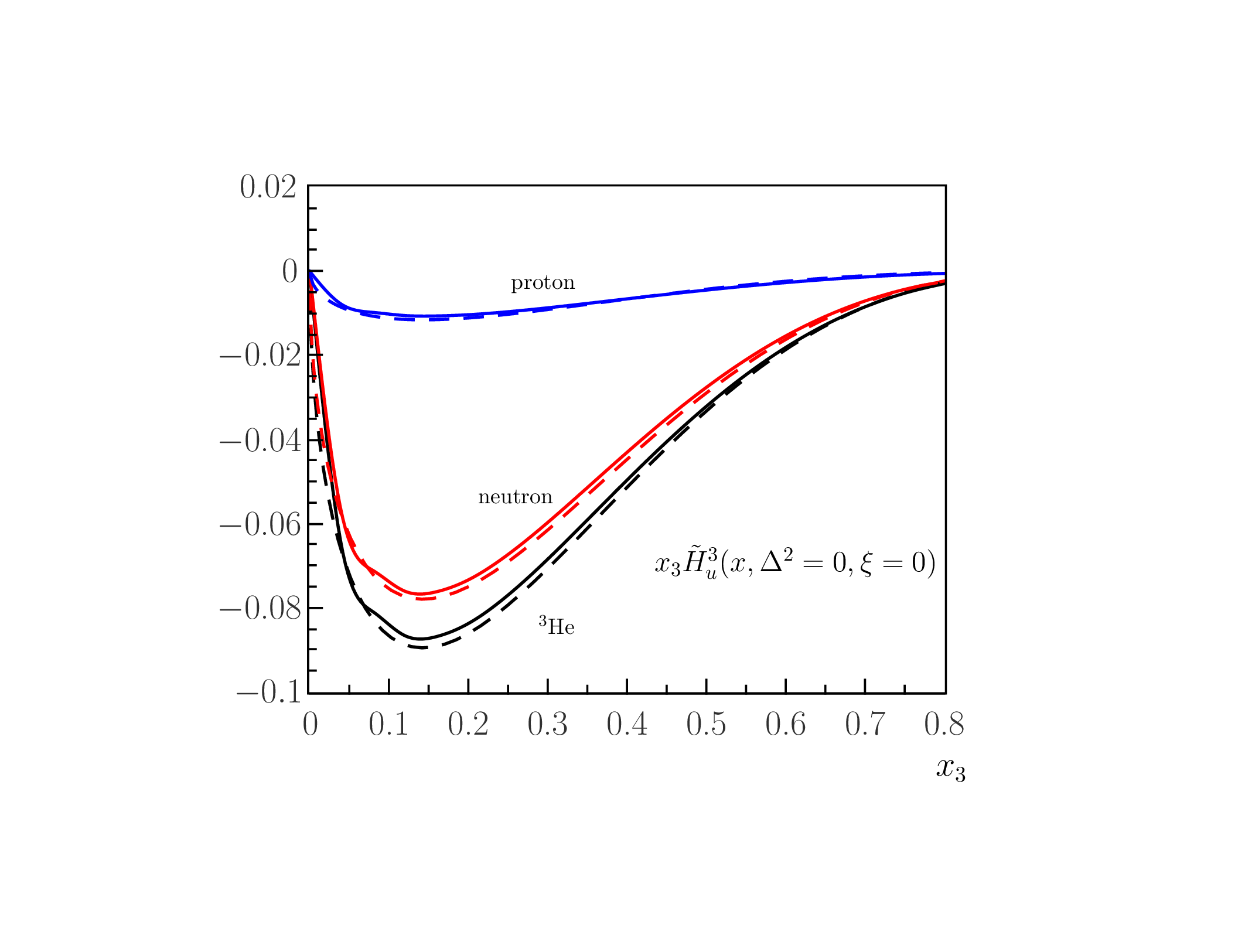}
\includegraphics{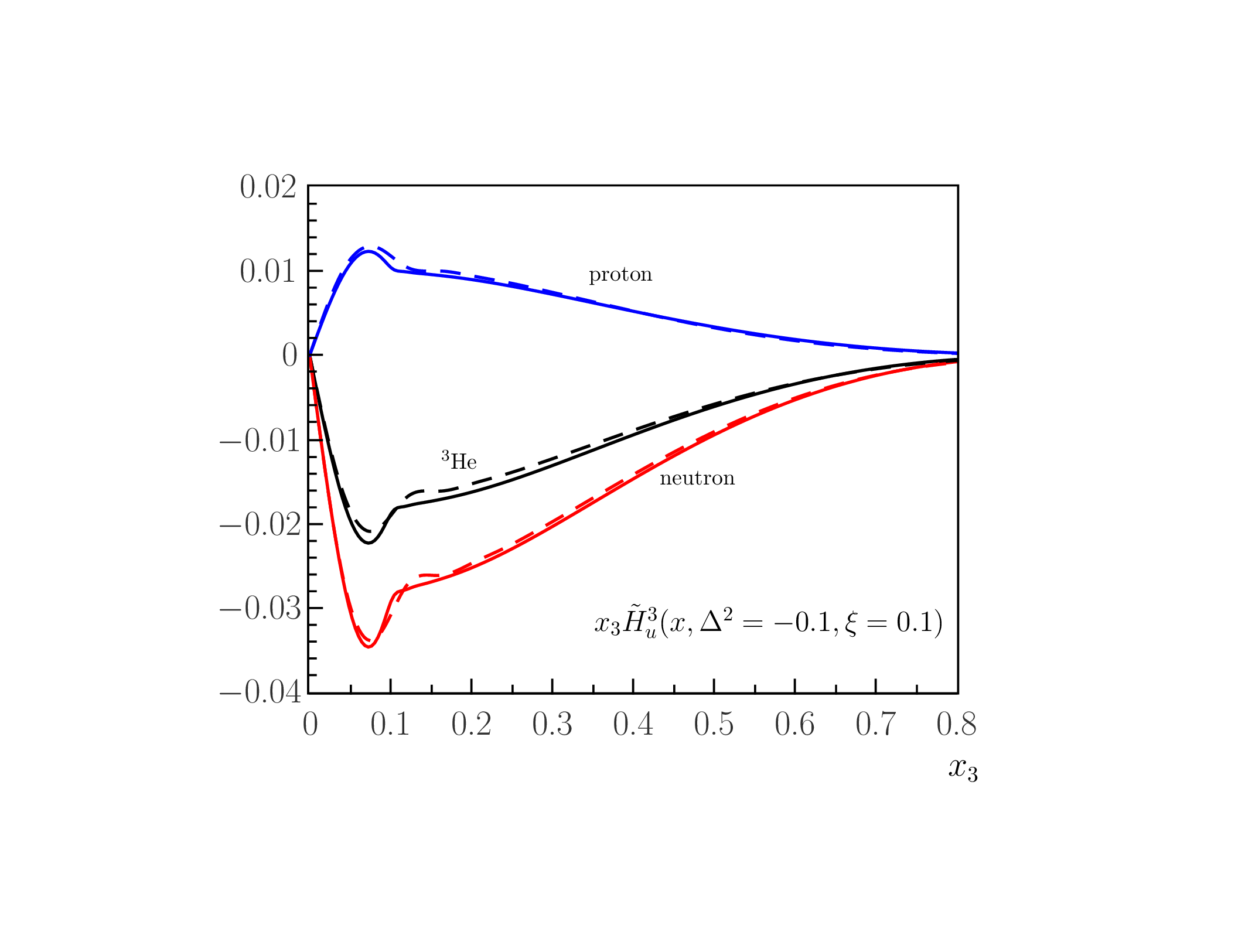}
%\end{figure*}
%
%\begin{figure*}[t]
%\vspace{7.5cm}
%\vskip-3.1cm
%\vskip-9.3cm
\caption{ 
The $q=u$ flavor GPD $x_3 \tilde{H}^{3,u}(x,\Delta^2,\xi)$, 
where $x_3 = (M_3/M) x$
and $\xi_3 = (M_3/M) \xi$, shown in the forward limit (left panel)
and at $\Delta^2 = -0.1
\ \mbox{GeV}^2$ and $\xi_3=0.1$ (right panel), 
together with the neutron and the proton
contribution. Solid lines correspond to the full IA result, Eq. (\ref{new}),
while the dashed ones represent the approximation
Eq. (\ref{sgu}).}
\end{figure*}

Deeply Virtual Compton Scattering, i.e. the process
$eH \longmapsto e'H' \gamma$ when  $Q^2\gg M^2$ ($Q^2=-q \cdot q$
is the transferred momentum 
between the leptons $e$ and 
$e'$, $\Delta^2$ the one between hadrons $H$ and $H'$ with
momenta $P$ and $P'$, and
$M$ is the nucleon mass), is one of
the cleanest processes to access GPDs.
The so called
skewedness, $\xi = - \Delta^+/(P^+ + P^{'+})$ 
\footnote{In this paper, $a^{\pm}=(a^0 \pm a^3)/\sqrt{2}$.}
is another relevant kinematical variable. 

%%%%%% NUCLEI
Despite serious difficulties to extract GPDs from  experiments, 
data for proton and nuclear targets are being analyzed,
see, i.e., Refs. \cite{data1,data2}.
The measurement of GPDs for nuclei
could be very useful to distinguish
between different mechanisms of nuclear
medium modifications of the nucleon structure,
an impossible task analyzing DIS experiments only
(this discussion started in Ref. \cite{deu}).
As usual, the neutron measurement, which requires
nuclear targets, is important. It permits in facts, 
together with the proton measurement,
a flavor decomposition of GPDs.
In studying observables related to 
the neutron polarization, $^3$He plays a special role, 
due to its spin structure (see, e.g., Refs. \cite{old,gdh}). 
This is valid also for GPDs studies.
Indeed, among the light nuclei
$^3$He is the only one for which the quantity
$\tilde{G}_M^{3,q}(x,\Delta^2,\xi) = H^{3}_q(x,\Delta^2,\xi)+
E^{3}_q(x,\Delta^2,\xi)$, i.e., the sum of its quark-helicity
independent GPDs $H_q$ and
$E_q$, could be dominated by the neutron
\cite{noi}:
in facts, the 
isoscalar targets
$^2$H and $^4$He are not suitable to this aim.
In Refs. \cite{noi}, it has been also shown 
to what extent this fact can be used to extract the neutron information.

The analysis of $^3$He GPDs 
in Impulse Approximation (IA) can be found in Refs. \cite{scopetta},
where, for the GPD $H$ of $^3$He,
$H_q^3$, 
convolution-like formulas
in terms of the corresponding
nucleon quantities has been derived.
Later, the treatment has been extended to
$\tilde{G}_M^{3,q}$
(see Ref. \cite{noi} for details),
and to the quark helicity flip GPD $\tilde{H}^3_q$
\cite{matserfb},
yielding
%\newpage
\begin{eqnarray}
 \tilde G_M^{3,q}(x,\Delta^2,\xi)  = 
\sum_N
\int dE 
\int d\vec{p}~
{\left [ P^N_{+-,+-}
%(\vec p,\vec p\,',E)  
-
P^N_{+-,-+} \right](\vec p,\vec p\,',E) }
{\xi' \over \xi}
\tilde G_M^{N,q}(x',\Delta^2,\xi'),
\end{eqnarray}
\vskip-3mm
\noindent
and
\begin{eqnarray}
{\tilde H_{q}^3(x,\Delta^2,\xi)}  = 
\sum_N
\int dE 
\int d\vec{p}
\,
{\left [ P^N_{++,++}
%(\vec p,\vec p\,',E)  
-
P^N_{++,--} \right](\vec p,\vec p\,',E) }
{\xi' \over \xi}
{\tilde H^{N}_q(x',\Delta^2,\xi')}~,
\label{new}
\end{eqnarray}
\vskip-3mm
\noindent
respectively.

In the last two equations, $x'$ and $\xi'$ are the variables 
for the GPDs of the bound nucleons, $p \, (p'= p + \Delta)$
is its 4-momentum in the initial (final) state and, eventually, 
proper components appear of the 
one body, off diagonal, spin dependent spectral function:

\begin{eqnarray}
 \label{spectral1}
 P^N_{SS',ss'}(\vec{p},\vec{p}\,',E) 
= 
\dfrac{1}{(2 \pi)^6} 
\dfrac{M\sqrt{ME}}{2} 
\int d\Omega _t
%\\
% \times   
\sum_{\substack{s_t}} \langle\vec{P'}S' | 
\vec{p}\,' s',\vec{t}s_t\rangle_N
\langle \vec{p}s,\vec{t}s_t|\vec{P}S\rangle_N~,
%\nonumber 
\end{eqnarray}
where $S,S'(s,s')$ are respectively the nuclear (nucleon) spin projections
in the initial (final) state,
and $E= E_{min} +E_R^*$, 
being $E^*_R$ the excitation energy 
of the interacting two-body recoiling system.
The main ingredient in the definition
Eq. (\ref{spectral1}) is
the intrinsic overlap integral
\begin{equation}
\langle \vec{p} ~s,\vec{t} ~s_t|\vec{P}S\rangle_N
=
\int d \vec{y} \, e^{i \vec{p} \cdot \vec{y}}
\langle \chi^{s}_N,
\Psi_t^{s_t}(\vec{x}) | \Psi_3^S(\vec{x}, \vec{y})
 \rangle~
\label{trueover}
\end{equation}
between the $^3$He wave function,
$\Psi_3^S$,  
and the final state, described by two different wave functions: 
{\sl i)}
the
eigenfunction $\Psi_t^{s_t}$, with eigenvalue
$E = E_{min}+E_R^*$, of the state $s_t$ of the intrinsic
Hamiltonian of the system of two {\sl interacting}
nucleons with relative momentum $\vec{t}$, 
which can be either
%a bound 
a deuteron nucleus 
or a scattering state, and 
{\sl ii)}
the plane wave describing 
the nucleon $N$ in IA.
For a numerical evaluation of Eqs. (1) and (2),
we used the overlaps Eq. (4) appearing in Eq. (3)
and corresponding to the analysis presented in Ref.
\cite{overlap} in terms of wave functions 
\cite{AV18}
evaluated using the
AV18  interaction \cite{pot}.

For the nucleonic $\tilde G_M^{N,q}$ 
and $\tilde H^{N}_q$, use has been made of a simple model
\cite{Rad1}, extended to evaluate
also spin dependent GPDs
(see Ref. \cite{noi,matserfb} for details).
%
%\vskip-7.01mm
Since $^3$He data are not yet available, 
one can verify only a few general 
GPDs properties, i.e., the forward limit and the first moments.
In particular the calculation of $H^{3}_q(x,\Delta^2,\xi)$ 
fulfills these constraints
\cite{scopetta}. 
Since there is 
no observable forward limit
for $E^{3}_q(x,\Delta^2,\xi)$,
for the $\tilde G_M^{3,q}(x,\Delta^2,\xi)$ calculation 
the only possible check is the first moment:
$
\sum_q \int dx \, \tilde G_M^{3,q}(x,\Delta^2,\xi) = G_M^3(\Delta^2);
$
where $G_M^3(\Delta^2)$ is the magnetic form factor (ff) 
of $^3$He.
The obtained result was found to be in agreement with previous
calculations (in particular, with the one-body part of the
AV18 calculation
presented in Ref. \cite{schiavilla}) and, 
for the low values of $\Delta^2$ which are relevant for the 
coherent process described here,
i.e., $-\Delta^2 \le 0.15$ GeV$^2$,
our results compare well also with the data.

As an example, let us discuss now, in a little more detail,
our calculation of 
$\tilde H^3_q$ \cite{matserfb}. 
We checked first of all that the forward limit of our expression,
Eq. (\ref{new}), reproduces formally and quantitatively
the results of  Ref. \cite{old} for polarized DIS
off $^3$He. On the other hand,
the first moment of $\tilde H^3_q$ is related
to the axial form factor of $^3$He, a poorly known observable
which does not permit therefore a direct check.
Having fulfilled at least the forward limit constraint,
we proceed now to analyze the proton and neutron
contributions to the $^3$He observable. 
The GPD $\tilde H_q^3$ is measured using a polarized target
and, as a consequence, it should
be dominated by the neutron contribution.
Let us show now to what extent
this property is obtained and how, thanks to this,
the neutron information could be extracted.

Results of the numerical evaluation of Eq. (\ref{new}) are
shown in Fig. 1. 
In the forward limit, 
the neutron contribution strongly dominates the
$^3$He quantity but, 
increasing $\Delta^2$, 
the proton contribution grows up,
in particular for the $u$ flavor
(see solid lines in Fig. 1).
A procedure to safely extract the neutron information from 
$^3$He
data is therefore necessary. This can be obtained by observing that
Eq. (\ref{new}) can be cast in the form 
\begin{eqnarray}
\tilde H^{3}_q(x,\Delta^2,\xi) \simeq   
\sum_N \int_{x_3}^{M_A \over M} { dz \over z}
h_N^3(z, \Delta^2, \xi ) 
\tilde H^{N}_q \left( {x \over z},
\Delta^2,
{\xi \over z},
\right)~,
%\nonumber
\end{eqnarray}
%\vskip-.9cm
where $h_N^3(z, \Delta^2, \xi )$ 
is a ``light-cone spin dependent off-forward momentum
distribution'', which
turns out to be strongly peaked around $z=1$,
close to the forward limit. 
Therefore, in this region,
for $x_3 = (M_A/M) x \leq 0.7$ one has:

\begin{eqnarray}
\tilde H^{3}_q(x,\Delta^2,\xi) 
& \simeq &   
\sum_N 
\tilde H^{N}_q \left( x, \Delta^2, {\xi } \right)
\int_0^{M_A \over M} { dz }
h_N^3(z, \Delta^2, \xi ) 
\nonumber
\\
& = &
G^{3,p,point}_A(\Delta^2) 
\tilde H^{p}_q(x, \Delta^2,\xi) 
+ 
G^{3,n,point}_A(\Delta^2) 
\tilde H^{n}_q(x,\Delta^2,\xi)~. 
%\nonumber
\label{sgu}
\end{eqnarray}
\vskip-1mm
In the equation above, we have introduced the axial point like ff, 
$G^{3,N,point}_A(\Delta^2)=\int_0^{M_A \over M} dz \, h_N^3(z,\Delta^2,\xi)$, 
which would give the nuclear axial ff if the proton and the neutron were 
point-like
particles.
These objects, at small values of $\Delta^2$, depend 
slowly on the potential used in the calculation, 
so that theoretical errors in their computation are small.
This can be realized observing that, in the forward limit, they reproduce
the ``effective polarizations'' of the protons ($p_p$) 
and the neutron ($p_n$) in $^3$He, whose values are rather similar 
if evaluated within different nucleon nucleon potentials 
(see Refs. \cite{old,overlap} for a comprehensive discussion). 
In particular, within the AV18 potential used here,
the values $p_n=0.878$ and $p_p= -0.024$ are obtained.
Then, Eq. (\ref{sgu}) can be used to extract the
neutron contribution from possible sets of data
for the proton and for $^3$He:
\vskip-5mm
\begin{eqnarray}
\label{extr}
\tilde H^{n,extr}(x, \Delta^2,\xi)  \simeq  
{1 \over G^{3,n,point}_A(\Delta^2)} 
\left\{ \tilde H^3(x, \Delta^2,\xi) 
 -  
G^{3,p,point}_{A}(\Delta^2) 
\tilde H^p(x, \Delta^2,\xi) \right\}~.
\end{eqnarray}

This comparison between the free neutron GPDs, used 
as input in the calculation,
and  the ones extracted using our calculation for
$\tilde H^3$ and the proton model for $\tilde H^p$, shows that the
procedure works nicely even beyond the forward limit. The only theoretical
input are axial point like ffs, 
which, as discussed above, are under good theoretical control.
The procedure works for $x \leq 0.7$,
the region where possible data could be gathered.
Analogous extraction procedures have been demonstrated to hold
for $\tilde H$ in ref. \cite{noi}.

In closing this section, we have shown that coherent DVCS off $^3$He at
low momentum transfer $\Delta^2$ as an ideal process to access 
the neutron GPDs;
if data were taken at higher $\Delta^2$, a relativistic treatment 
\cite{ema}
and/or the inclusion of many body currents, beyond the present IA scheme, 
should be implemented. 
The next step of this investigation will be the evaluation of cross section
asymmetries relevant to DVCS experiments at JLab kinematics, using
the obtained GPDs $H, E$ and $\tilde H$ of $^3$He as a theoretical
input.
At the beginning, the leading twist analysis of DVCS for a spin 1/2 target,
presented in Ref. \cite{diet}, will be performed.  

\section{Extraction of neutron asymmetries from SIDIS experiments off $^3He$}

\label{sidis}

As discussed in the Introduction, information on the three-dimensional 
structure of the nucleon in
momentum space  
can be obtained studying TMDs and transverse momentum
dependent fragmentation functions (TMFFs) in SIDIS processes.
The Sivers TMD \cite{siv} and the Collins TMFF \cite{col}, describing  
correlations between the spin and the momentum of a parton,
are due to leading-twist final state interactions at
the parton level and
have therefore a very interesting
dynamical content \cite{brod}.
The Sivers and Collins functions are  
extracted from single spin asymmetries
(SSAs) built from 
differential cross sections of SIDIS of unpolarized electrons
off transversely polarized targets.
The present data for the processes
${\vece p}(e,e'\pi)x$ \cite{due1} 
and $\vece D(e,e'\pi)x$ \cite{tre} show a 
strong flavor dependence and
measurements 
with a $^3\vece{\mathrm{He}}$ target 
have been proposed and performed
(see Ref.  \cite{tre1}
for the first data at JLab with a 6 GeV electron beam). 
Further accurate  experiments  are 
planned at the {12-GeV} upgrade of JLab \cite{quattro}. 
As in the GPDs case discussed in the previous section,
neutron data are important to achieve the flavor 
separation of the Sivers and Collins distributions \cite{cinque},
and polarized $^3$He targets play a special role. 
To obtain a reliable information one has to take into account 
the nuclear structure of $^3\mathrm{He}$ as carefully as possible.
Initially, the process can be described 
using the Plane Wave Impulse Approximation (PWIA),
leading to dynamical nuclear effects described 
by the spin-dependent spectral function of 
$^3$He, ${\rm P}_{\sigma,\sigma^\prime} (\vec p, E)$, 
(see, e.g.  \cite{Ciofi1})
that yields
the probability distribution to find a nucleon with given removal energy 
{(as a matter of facts, the spectator pair is interacting)},
three-momentum and polarization inside the nucleus.
By using  this formalism, it has been shown long time ago that,
in spin dependent DIS,
one can safely extract 
the  neutron longitudinal asymmetry, $A_n $,
from the corresponding $^3$He  observable, $A_3^{exp}$ \cite{old}. 
One obtains
%  \vskip -0.4cm
\be
  {{A_n }}\simeq  \left 
( {A^{exp}_3} - 2 
{p_p} f_p
{{A^{exp}_p}} \right )/(p_n f_n )\label{dis}\ee
with  $p_{n(p)}$  the
neutron (proton) effective polarization inside the polarized $^3$He, and
$f_{n(p)}$,  the dilution factor. Values
of $p_n$ and $p_p$ obtained using the AV18 interaction \cite{pot} are  
$ {{p_p}} = -0.024  $, 
${{p_n}}= 0.878 $  (see, e.g., \cite{Sco}).
In Ref. \cite{Sco}, an analogous extraction was demonstrated
to work also in SIDIS,
if the Bjorken limit and PWIA are assumed, and 
applied to the SSA of a
transversely polarized $^3\vece{{\rm He}}$ target, obtained 
from the  process $^3\vece{{\rm He}}( e,e'\pi)X$, in order to obtain the 
SSA of a transversely polarized neutron. 
In that approach, the SSAs of  $^3\vece{\rm { He}}$
can be expressed as  convolutions of  
$ {\rm P}_{\sigma,\sigma^\prime} (\vec p, E)$ and proper
combinations of
suitable 
TMDs and TMFFs. 
The same extraction procedure has been also applied in combination 
with a Monte Carlo 
simulating the kinematics of the experiment E12-09-018 \cite{sei}.
Indeed
Eq. (\ref{dis}) has been used by the 
JLab Hall A Collaboration to extract, for the first time, 
the Collins and Sivers moments from a transversely polarized 
$^3\mathrm{He}$ target \cite{tre1}.
Since  existing measurements are limited in statistical accuracy and 
kinematics coverage,  
an extensive program of high precision measurements of SIDIS off 
$^3\mathrm{He}$ will be part of the JLab program at 12 GeV \cite{quattro}. 
The expected statistical accuracy is of the order of percent in a 
wide range of multi-dimensional kinematical binning; for this reason the 
PWIA could be no longer sufficient and Final State Interactions (FSI)
between the detected meson and the remnants of the process, 
not considered in PWIA, may have a relevant role.
Let us discuss now a proper way to introduce this mechanism, leading
to a
distorted spin-dependent spectral function of the $^3\mathrm{He}$.   
\begin{figure}
%{l}{0.3\textwidth} 
%\vspace{-20pt}
\includegraphics[width=10cm]{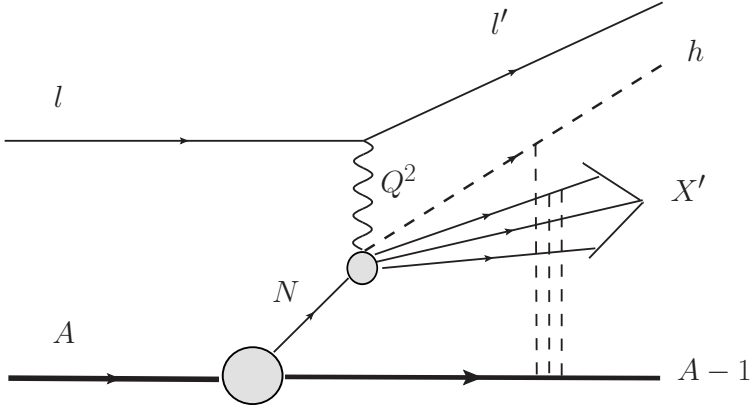}
 \caption{Final State Interaction between the fully interacting
$(A-1)$ spectator system and the debris
produced by the
absorption of a virtual photon by a nucleon in the nucleus.}
%\vspace{-20pt}
\label{fig:1}
\end{figure}
The JLab SIDIS experiments will exploit an electron beam at 8.8 and 11 GeV 
off $^3\mathrm{He}$ polarized gaseous target; the relative energy 
between the $(A-1)$ system and the system of the detected pion and the 
remnant (depicted in Fig. \ref{fig:1}) is a few GeV
and therefore FSI can be treated 
within a generalized eikonal approximation (GEA). 
This framework was already successfully applied to unpolarized 
SIDIS \cite{sette}, 
and in a recent paper the {\it distorted} spin-dependent spectral function 
has been calculated for the so-called {\it spectator} SIDIS, 
where a slow $(A-1)$ 
nucleon system, acting as a spectator of the photon-nucleon interaction, 
is detected, while the produced fast hadron is not \cite{otto}.
In the following we discuss standard SIDIS, where all the possible state of 
the two-nucleon spectator system have to be considered. 

The {\it distorted} spin-dependent spectral function for a polarized $^3\mathrm{He}$ target can be written as 
\be
S^{N \,{\bf S}_3}_{\lambda\lambda'}(E,{\bf p}_{mis}) = 
\sum_{f_2} 
\sum \! \!\! \!\! \!\! \!\int_{~\epsilon^*_2}\rho\left(
\epsilon^*_2\right)\,
{ \tilde {\cal O}}_{\lambda\lambda'}^{N \, {\bf S}_3 \, f_2}
(~\epsilon^*_2,{\bf p}_{mis})
\,
\delta\left( { E+ M_3-m_N-M^*_2}\right)
\ee
with the product of distorted overlaps, given in terms of the intrinsic
Jacobi coordinates 
${\bf {r}}$
and $ {{\brho}}$, defined by
\begin{eqnarray}
{\tilde {\cal O}}_{\lambda\lambda'}^{N \, {\bf S}_3 \, f_2}
(\epsilon^*_2,{\bf p}_{mis}) & = & 
\la \lambda,
\phi_{\epsilon_2^*}^{f_2}({\bf r})
e^{-i{\bf p}_{mis}{{\brho}}} {\cal G}({\bf{r}},{{\brho}})
|
\Psi_3^{{\bf S}_3}({\bf{r}},{{\brho}})\ra
\nonumber \\ 
& \times &
\la \Psi_3^{{\bf S}_3}({\bf{r}}',{{\brho}}')| \lambda', 
{\cal G}({\bf{r}}',{{\brho}}')
\phi_{\epsilon_2^*}
^{f_2}
({\bf{r}}')
e^{-i{\bf p}_{mis}{{\brho}}'}\ra.
\label{overfsi}
\end{eqnarray}    
{where i) $\rho\left(\epsilon^*_{2}\right)$ is the density of the spectator 
pair  with intrinsic 
energy $\epsilon^*_{2}$, ii) $|\Psi_3^{{\bf S}_3}({\bf{r}},{{\brho}})\ra$ 
is the ground state of the  3-nucleon system
with polarization ${\bf S}_3$} iii) $E$ is the removal energy 
$E = \epsilon^*_{2} + B_3$, 
${\bf p}_{mis}$ is the three momentum of the spectator 
pair. 

\begin{figure}
\includegraphics[width=\textwidth]{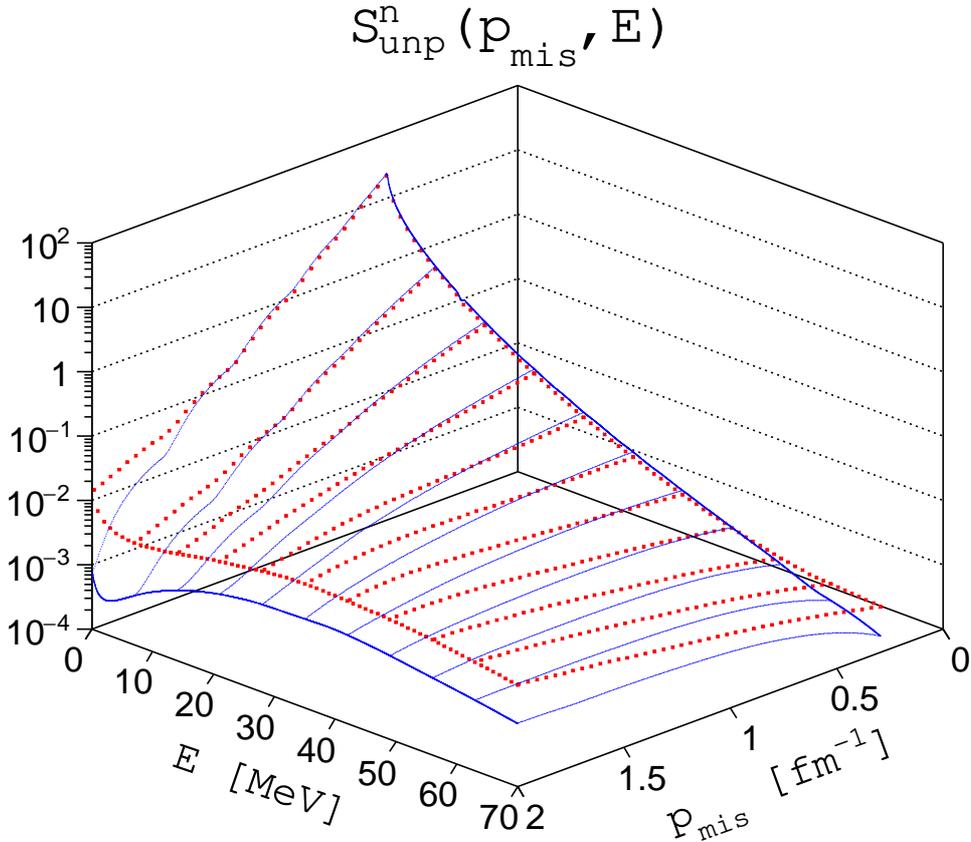}
  \centering
  \caption{The $^3$He spectral function, for the neutron,
in the unpolarized case, as a function of $p_{mis}=|{\bf p}_{mis}|$ and
of the removal energy $E$, in PWIA (full lines) and with
FSI taken into account in the present GEA framework (dotted lines).
The spectral functions are shown for the
values of $E$ and $p_{mis}$ that contribute to the calculus
of the SIDIS cross section 
when $A(p_N\cdot q)/(P_3\cdot q) = 0.86$ 
at ${\cal E}=$ 11 GeV and
$Q^2 \simeq 7.6$ (GeV/c)$^2$ (see 
Ref. \cite{dieci} for more details).}
\label{fig:2}
\end{figure}

The Glauber operator in Cartesian coordinates is given by 
\be
{\cal G}({\bf{r}}_1,{\bf{r}}_2,{\bf{r}}_3)={\cal G}({\bf{r}},{{\brho}})=\prod\limits_{i=2,3}
\left[ 1-
\theta({\bf{r}}_{i||}-{\bf{r}}_{1||})
\Gamma \left( {\bf{r}}_{i\perp}-{\bf{r}}_{1\perp},{\bf{r}}_{i||}-{\bf{r}}_{1||}
\right) \right ]~,
\label{gl}
\ee
where $\hat{\bf{r}}_{i\perp}$ and  $\hat{\bf{r}}_{i||}$ are the perpendicular and the 
parallel components of ${\bf{r}}_i$ with respect to 
the direction of the debris. The profile function 
\be
\Gamma
(
{\bf{r}}_{i\perp}
-
{\bf{r}}_{1\perp}
,
{\bf{r}}_{i||}
-
{\bf{r}}_{1||}
)
\, =\,
\frac{
(1-i\,\eta)\,\,
\sigma_{eff}
(
{\bf{r}}_{i||}
-
{\bf{r}}_{1||}
)
}
{4\,\pi\,b_0^2}\,
\exp 
\left[ -
\frac{
(
{\bf{r}}_{i\perp}
-
{\bf{r}}_{1\perp}
)^2
}
{2\,b_0^2}
\right]~~~,
\label{eikonal}
\ee
unlike that occurring in
the standard Glauber approach, depends not only on the impact parameter 
but also on the longitudinal 
separation through an effective cross section, $\sigma_{eff}$. 
For the latter we used the model proposed
and exploited in previous works 
{(for details on the model  see  Ref. \cite{sette,otto})}.
To recover the PWIA formulation, one has simply to put ${\cal G}\equiv 1$
in Eq. (11). 
In Fig. \ref{fig:2}, a plot of the 
$^3\mathrm{He}$ spectral function,
evaluated within plane wave impulse approximation
or taking FSI into account, is shown,
for the neutron, 
in the unpolarized case.  

\begin{figure}[h]
\includegraphics[scale=0.54]{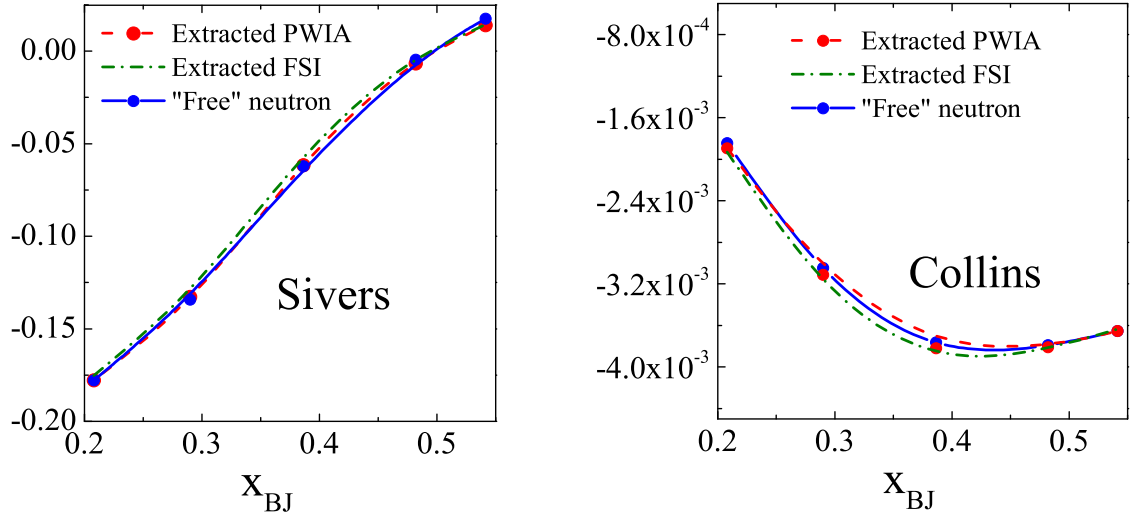}
\caption{Neutron asymmetries extracted through Eq. (8) from the Sivers 
(Left panel) and Collins (Right panel) $^3$He asymmetries, 
with and without FSI
taken into account
in the actual kinematics of JLab at 12 GeV \cite{quattro}. 
Preliminary results to appear in \cite{dieci}.}
\label{fig:3}
\end{figure}
The relevant part in the extraction of the transversely polarized neutron's 
SSAs is the transverse spectral function, given by
%\be
%{\cal S}^{N \, \perp}(E,{\bf p}_{mis})={\cal S}^{N \, \frac12 -\frac12}_{\frac12 -\frac12}(E,{\bf p}_{mis})
%+ {\cal S}^{N \, -\frac12 \frac12}_{\frac12 -\frac12}(E,{\bf p}_{mis})
%\label{trspectr}
%\ee
a proper combination of non-diagonal elements of the spin-dependent
spectral function defined in Eq. (9).
In general, the transverse 
spin-dependent SF evaluated in PWIA,
${\cal S}^{\perp (PWIA)}$, and the corresponding distorted
quantity,
${\cal S}^{\perp (FSI)}$,
can be quite different and the corresponding
effective polarizations can differ
by 10-15\%. Nevertheless, in Eq. (\ref{dis}) the effective polarizations 
occur in products with the dilution factors and to a large 
extent it turns out that, in the kinematical range explored at JLab, 
$p_p^{PWIA}f_p^{PWIA}\approx p_p^{FSI}f_p^{FSI}$ ,  
$p_n^{PWIA}f_n^{PWIA}\approx p_n^{FSI}f_n^{FSI}$ \cite{best}. This allows
one to safely adopt the usual extraction, as shown in Fig. \ref{fig:3}, 
and therefore the goal of a sound flavor
decomposition of TMDs seems definitely safe.    

\section{Conclusions and Perspectives}

Let us summarize the status of our calculations of DVCS and SIDIS processes
with $^3$He targets.
In section 2, we have shown realistic calculations of GPDs of $^3$He,
in plane wave impulse approximation, the essential ingredients
for the evaluation of DVCS cross sections.
Although DVCS off $^3$He turns out to be promising for the extraction
of neutron GPDs and relevant to understand the $A-$dependence
of the modification of the parton structure of bound nucleons,
for technical reasons, 
the present experimental program of JLab at 12 GeV will collect
data for DVCS off deuteron and $^4$He targets.
Preliminary data of DVCS off $^4$He, gathered
with JLab at 6 GeV, are already available
and should be published soon \cite{moam}.
They deserve the attention of our community, which
could perform precise calculations.
Anyway, as explained in section 2, isoscalar targets 
such as deuteron or $^4$He are not
useful for a complete flavor separation of GPDs. 
Besides, a scalar nucleus as $^4$He cannot be used 
for the measurement of helicity dependent GPDs, and the
same quantity is hardly extracted from deuteron data.
At the planned 
Electron Ion Collider \cite{eic}, where slow nuclear recoiling systems
will be easily detected and polarized nuclear beams will be naturally
available, the proposal of polarized DVCS off $^3$He
and the measurement of the neutron helicity dependent GPD, 
$\tilde H$, using the technique
described in Section 2, should become feasible.

As described in this talk for both DVCS and SIDIS,
the standard theoretical description
of few-nucleon systems, where nucleon and pion degrees of freedom are taken 
into account, has achieved a very high degree of sophistication.
Nonetheless, 
one should try to fulfill, as much as possible, the relativistic 
constraints, dictated by covariance with respect to the Poincar\'e group, 
if processes involving nucleons with large 3-momentum are considered and 
a high precision is needed. 
At least, one should carefully deal with the boosts of the nuclear states, 
$|\Psi_{init}\rangle$ and $|\Psi_{fin}\rangle$.
In particular a relativistic treatment is important to precisely 
describe the JLab program @ 12 GeV for few-body systems (see, 
e.g., Refs. \cite{DIS}, \cite{SIDIS}).
To this aim,
in the last few years, we have developed 
a relativistic description of $^3$He
using a Poincar\'e covariant spectral function built up within 
the light-front Hamiltonian dynamics  (LFHD) \cite{prclast}.

Indeed, the Relativistic Hamiltonian Dynamics (RHD) of an interacting 
system with a fixed number of on-mass-shell constituents 
(see, e.g., \cite{KP}), 
associated to the Bakamijan-Thomas construction of the Poincar\'e generators 
\cite{Baka}, permits a fully Poincar\'e covariant description of DIS, 
SIDIS and DVCS off $^3He$. The light-front (LF) form of RHD has been adopted 
\cite{KP}, which has 
a subgroup structure of the LF boosts (with a separation of the intrinsic 
motion from the global one, which is very important for our aim) 
and a meaningful Fock 
expansion (see, e.g., Ref. \cite{Brodsky}).
Furthermore, within the LFHD, one can take advantage of the successful 
non-relativistic phenomenology that has been 
developed for the nuclear interaction.

The LF spin-dependent spectral function obtained from the LF wave functions 
for the two- and the three-nucleon systems has been defined
in \cite{prclast,best}.
With respect to previous attempts, the essential difference 
is the definition of the nucleon momentum: in our approach, it is 
the intrinsic momentum of the nucleon in the cluster 
formed by the nucleon 
and by the fully interacting (A-1) spectator system with given mass. 
This definition implements the macrocausality: if a system is separated 
into disjoint subsystems by a sufficiently large spacelike separation, 
then the subsystems behave as independent systems.
The proposed formalism can find useful applications in deep 
inelastic scattering, since the LF momentum distribution fulfills both 
normalization and momentum sum rules, while the cluster separability 
introduces new binding effects with respect to previous approaches.

We are presently applying our relativistic spectral function 
to study the role of relativity in
the EMC effect for 
$^3$He. Preliminary results have been presented
in Ref. \cite{lf15}. 
We plan now to complete the LF analysis of the EMC effect
for $^3$He. 
As next steps, we will calculate a LF non-diagonal spectral function,
to analyze DVCS cross sections, where relativity could
play an essential role, if high transfer momentum is involved,
as it happens for hadrons 
electromagnetic form factors (see, e.g., Ref. \cite{ff}).
Eventually,
FSI should be added to our LF relativistic description of $^3\mathrm{He}$,
to study SIDIS processes and calculate single spin asymmetries. 

\section*{Aknowledgmenys}
S.S. thanks the organizers of the Conference
``Theoretical Nuclear Physics in Italy, 2016'' for the invitation
to give this talk in a lively and timely event.

\end{document}